\documentclass[10pt,conference]{IEEEtran}
\usepackage{amsmath,amssymb,amsfonts}
\usepackage{algorithmic}
\usepackage{graphicx}
\usepackage{textcomp}
\usepackage{xcolor}
\usepackage{xspace}
\usepackage{tabularx}
\usepackage{hyperref}
\usepackage{balance}
\usepackage{xcolor}
\usepackage{colortbl}

\newcommand{\etal}{et al.\xspace}
\newcommand{\ie}{i.e.\xspace}
\newcommand{\eg}{e.g.\xspace}
\newcommand{\fig}[1]{Fig.~\ref{#1}}

\newcommand{\figsize}{0.95\columnwidth}
\newcommand{\debian}{Debian\xspace}
\newcommand{\stable}{Debian Stable\xspace}

\newcommand{\testing}{Debian Testing\xspace}
\newcommand{\unstable}{Debian Unstable\xspace}
\newcommand{\ubuntu}{Ubuntu\xspace}
\newcommand{\fedora}{Fedora\xspace}
\newcommand{\centos}{CentOS\xspace}
\newcommand{\centoslatest}{CentOS 8\xspace}
\newcommand{\centosseven}{CentOS 7\xspace}
\newcommand{\arch}{Arch Linux\xspace}
\newcommand{\linux}{Linux\xspace}
\definecolor{visibleBlue}{RGB}{0,160,225}

\usepackage[disable]{todonotes}

\usepackage{mdframed}
\newenvironment{custombox}{\begin{mdframed}[nobreak=true, innerleftmargin=5pt, innerrightmargin=5pt, innertopmargin=2.5pt, innerbottommargin=2.5pt]}{\end{mdframed}}

\usepackage{tikz}
\usetikzlibrary{arrows,decorations.markings}
\usetikzlibrary{matrix,positioning}
\usetikzlibrary{fit,automata,positioning,calc}
\usetikzlibrary{graphs}
\usetikzlibrary{graphs.standard}
\usetikzlibrary{shapes.geometric}
\usetikzlibrary{snakes}
\usetikzlibrary{arrows.meta}
\usetikzlibrary{decorations}
\pgfdeclaredecoration{discontinuity}{start}{
	\state{start}[width=0.5\pgfdecoratedinputsegmentremainingdistance-0.5\pgfdecorationsegmentlength,next state=first wave]
	{}
	\state{first wave}[width=\pgfdecorationsegmentlength, next state=second wave]
	{
		\pgfpathlineto{\pgfpointorigin}
		\pgfpathmoveto{\pgfqpoint{0pt}{\pgfdecorationsegmentamplitude}}
		\pgfpathcurveto
		{\pgfpoint{-0.25*\pgfmetadecorationsegmentlength}{0.75\pgfdecorationsegmentamplitude}}
		{\pgfpoint{-0.25*\pgfmetadecorationsegmentlength}{0.25\pgfdecorationsegmentamplitude}}
		{\pgfpoint{0pt}{0pt}}
		\pgfpathcurveto
		{\pgfpoint{0.25*\pgfmetadecorationsegmentlength}{-0.25\pgfdecorationsegmentamplitude}}
		{\pgfpoint{0.25*\pgfmetadecorationsegmentlength}{-0.75\pgfdecorationsegmentamplitude}}
		{\pgfpoint{0pt}{-\pgfdecorationsegmentamplitude}}
	}
	\state{second wave}[width=0pt, next state=do nothing]
	{
		\pgfpathmoveto{\pgfqpoint{0pt}{\pgfdecorationsegmentamplitude}}
		\pgfpathcurveto
		{\pgfpoint{-0.25*\pgfmetadecorationsegmentlength}{0.75\pgfdecorationsegmentamplitude}}
		{\pgfpoint{-0.25*\pgfmetadecorationsegmentlength}{0.25\pgfdecorationsegmentamplitude}}
		{\pgfpoint{0pt}{0pt}}
		\pgfpathcurveto
		{\pgfpoint{0.25*\pgfmetadecorationsegmentlength}{-0.25\pgfdecorationsegmentamplitude}}
		{\pgfpoint{0.25*\pgfmetadecorationsegmentlength}{-0.75\pgfdecorationsegmentamplitude}}
		{\pgfpoint{0pt}{-\pgfdecorationsegmentamplitude}}
		\pgfpathmoveto{\pgfpointorigin}
	}
	\state{do nothing}[width=\pgfdecorationsegmentlength,next state=do nothing]{
		\pgfpathlineto{\pgfpointdecoratedinputsegmentlast}
	}
	\state{final}
	{
		\pgfpathlineto{\pgfpointdecoratedpathlast}
	}
}

\def\BibTeX{{\rm B\kern-.05em{\sc i\kern-.025em b}\kern-.08em
		T\kern-.1667em\lower.7ex\hbox{E}\kern-.125emX}}

\begin{document}
\title{A Quantitative Assessment of \\ Package Freshness in Linux Distributions\\
	\thanks{Fonds de la Recherche Scientifique -- FNRS}
}

\author{\IEEEauthorblockN{Damien Legay}
	\IEEEauthorblockA{\textit{Software Engineering Lab} \\
		\textit{University of Mons}\\
		Mons, Belgium \\
		damien.legay@umons.ac.be \\
		ORCID 0000-0001-6811-6585}
	\and
	\IEEEauthorblockN{Alexandre Decan}
	\IEEEauthorblockA{\textit{Software Engineering Lab} \\
		\textit{University of Mons}\\
		Mons, Belgium \\
		alexandre.decan@umons.ac.be \\
		ORCID 0000-0002-5824-5823}
	\and
	\IEEEauthorblockN{Tom Mens}
	\IEEEauthorblockA{\textit{Software Engineering Lab} \\
		\textit{University of Mons}\\
		Mons, Belgium \\
		tom.mens@umons.ac.be \\
		ORCID 0000-0003-3636-5020}
}

\pagestyle{plain}
\maketitle

\begin{abstract}
	\linux users expect fresh packages in the official repositories of their distributions. Yet, due to philosophical divergences, the packages available in various distributions do not all have the same degree of freshness. Users therefore need to be informed as to those differences.
	Through quantitative empirical analyses, we assess and compare the freshness of 890 common packages in six mainstream \linux distributions.
	We find that at least one out of ten packages is outdated, but the proportion of outdated packages varies greatly between these distributions.
	Using the metrics of update delay and time lag, we find that the majority of packages are using versions less than 3 months behind the upstream in 5 of those 6 distributions.
	We contrast the user perception of package freshness with our analyses and order the considered distributions in terms of package freshness to help \linux users in choosing a distribution that most fits their needs and expectations.
\end{abstract}

\section{Introduction}
Since its inception in 1991, the \linux operating system has continued to expand and evolve.
Its kernel has grown at a superlinear rate~\cite{Godfrey2000,robles2005evolution}. Since Linux uses the very permissive open source license GNU GPL, it has been forked into a myriad of variants, called \textit{distributions}. These distributions are often built around a \textit{package manager} that complements the \linux kernel with a plethora of third-party software \textit{packages}.
Much of the functionality provided by a distribution to its end-users comes from these third-party packages.
They are provided to users via distribution-specific \textit{software repositories}. As a result, \linux distributions form \textit{software ecosystems} in which packages interact in complex relationships of dependency, co-installability~\cite{Vouillon2011}, redundancy and complementarity.
As such, the versions of packages available in the official repositories (repositories that are enabled by default) of different distributions may vary according to the philosophy adopted by the creators and maintainers of the distribution, as they take different approaches to nurturing the health of their distributions.

Some maintainers place great emphasis on maximising the \textit{stability} of the distribution, to avoid the risk of introducing changes that could potentially introduce package incompatibilities or break prior functionality. A distribution known to markedly emphasise stability is the Stable branch of \debian. 
Some maintainers put security concerns at the forefront, aiming to make their distributions as resistant as possible to nefarious acts. One such distribution is Qubes OS~\cite{rutkowska2010qubes}, which minimises the potential impact of security vulnerabilities by isolating software components as much as possible through the use of virtual machines.
Other maintainers, such as those of \arch, prioritise package \textit{freshness} by endeavouring to incorporate package updates as quickly as possible.

In prior work~\cite{legay2020ICSME}, we conducted a qualitative survey of 170 \linux users revealing that \linux users consider package freshness important, as package updates are a source of security patches, bug fixes and new features. This is also attested by the existence of package freshness monitoring services such as \textsf{Repology} and \textsf{DistroWatch}.
The survey also highlighted that users tend to rely on their distributions' official repositories to install and update packages.
Therefore, it is important for users to be well-informed as to the relative freshness of the packages in various \linux distributions.
The survey examined the user perception of package freshness in the official repositories of the distributions, observing vast discrepancies between distributions, from users of Arch expecting package updates to be deployed within days, to users of \stable and \centos deeming that it would take months.
The survey also established that whenever fresh versions of packages are not available within the official repositories of the distribution, users have to use other means of updating, which can be detrimental to their system's stability and security.
Therefore, in order to evaluate to which extent user perception matches objective quantitative evidence and to enable them to make an informed choice of distribution, we empirically measure and compare the package freshness of six popular distributions: \arch, \centos, \stable, \unstable, \fedora and \ubuntu, over a period of 5 years.
We do so by studying three research questions for each distribution.
$RQ_1$: How prevalent are outdated package versions in \linux distributions?
$RQ_2$: How long have those package versions been outdated?
$RQ_3$: To which extent are deployed packages outdated?

\section{Related Work}
The notion of freshness has been studied in non-\linux environments.
Cox \etal~\cite{Cox2015} proposed system-level metrics to quantify a software system's \textit{dependency freshness}, that is to say how up-to-date the system's dependencies are. 
Gonzalez-Barahona \etal~\cite{gonzalez2017technical} introduced the concept of {\em technical lag} as a measure of how outdated a system is with respect to its dependencies. They defined the technical lag in terms of a lag function and lag aggregation function for packages. This notion was generalised further and used by Zerouali \etal~\cite{zerouali2019formal}. 

Aspects of package co-installability and package dependency in \linux have been well studied, particularly with regards to \debian.
Vouillon \etal~\cite{Vouillon2011,Vouillon2013} and Claes \etal~\cite{ClaesEtAl2015Debian} examined the co-installability problem as it applied to the \debian ecosystem and how it could be solved. Similarly, Artho \etal~\cite{ArthoSCTZ12} proposed detection and prevention strategies for this problem.
Galindo \etal~\cite{galindo2010debian} characterised the relationships between packages within \debian, proposing a language to describe whether a package depends on another, has co-installability issues with another or provides the same functionality or a superset of the functionality of another.
Nguyen and Holt~\cite{nguyen2012life} studied the life cycle of \debian packages. They compared the age of packages in Debian Stable, Unstable and Testing, defining package age as the time delta between its introduction into the distribution and its removal from \debian or its update to a newer version.

There has been little focus on the freshness of packages in \linux distributions.
Gonzalez-Barahona \etal~\cite{gonzalez2009macro} observed in 2009 that one out of eight packages within \stable (12\%) was not updated at all during a nine-year timespan, from \stable 2.0 (released on 1998-07-24) to 4.0 (released on 2007-04-08).
The work most closely related to this paper is a 2009 Bachelor's thesis by Shawcroft~\cite{shawcroft2009open} comparing the freshness of packages in 8 \linux distributions (\arch, \debian, Gentoo, \fedora, OpenSUSE, Sabayon, Slackware and \ubuntu).
The thesis reports that around 20\% of the packages in \arch are outdated and from 40\% to 60\% in the other distributions, and that \debian and Slackware packages were, on average, more obsolete than those of the other distributions. Although his analysis concerns 8 \linux distributions, it only covers 137 packages.

\section{Methodology}
\label{methodology}
This section presents the methodology for our empirical study, for which a replication package is available on Zenodo\footnote{\url{http://doi.org/10.5281/zenodo.4446468}}.

In order to study package freshness in \linux distributions, we need to select a set of relevant distributions. According to \textsf{Distrowatch}, there are 913 \linux distributions, 274 of which are considered ``active''.
Not all distributions fall within the scope of this study (\eg Bicom System is a distribution whose sole purpose is to serve as a telephony platform).
We focus on general-purpose GNU-based distributions, and select \arch, \debian, \centos, \fedora and \ubuntu. These distributions were found to be used by 81\% of respondents in our survey of Linux users~\cite{legay2020ICSME}.

These distributions have different release policies. Most of them use point releases, in which a new version of the distribution is released at regular intervals. \ubuntu has a fixed release cycle with six-month intervals, releasing in April and October.
\fedora also releases two versions a year, but on a looser schedule. 
\centos is based on the source code of Red Hat Entreprise Linux (RHEL). 
For example, \centos 6.5 is based on the fifth update of the sixth release of RHEL. 
\arch follows a rolling release policy, wherein packages are constantly updated. There are therefore no explicit version releases of \arch.
\debian includes several distributions. \stable is the officially recommended distribution.
\testing and \unstable are development branches: packages updates and new packages start in the \unstable rolling release and are continuously updated. When a given package or update fulfils certain requirements, it is moved to \testing. Every 18 months, Testing is frozen, only receiving critical fixes from Unstable. Six months later, a new release is made: Testing becomes Stable and is unfrozen. The previous \stable becomes Debian Oldstable.

To conduct empirical analyses on the freshness of packages in the selected distributions, we require data on the package versions contained within them.
We relied on the repositories and archives of official distributions to obtain this data. For distributions relying on point releases, we gathered data on the package versions present in the distribution at the moment of release. For distributions relying on rolling releases, we gathered data on daily snapshots of the distributions. We focused the analysis on a five-year observation period: $[2015, 2020[$.
We therefore selected the distribution releases corresponding to that period. They are, respectively: \fedora 23 to 31, \ubuntu 15.04 to 19.10 and \centos 7.1 to 7.7. We selected  \centosseven over \centoslatest for the analysis because \centoslatest only had one release during the observation period (8.0, released on 2019-09-24). For \debian, we included Stable 8 to 10 and daily snapshots of Unstable. We did not include Testing because, at the moment of release, it is identical to Stable.

To be able to compare distributions, we needed to select a set of packages that are common to all distributions, \ie packages that are present in at least one snapshot of each distribution. Yet, different distributions might not adopt a package under the same name.
For example, Xephyr X-server is available in \debian-derived distributions as the {\sf xserver-xephyr} package and as {\sf xorg-server-xephyr} in \arch.
To take into account these package name variations across distributions, we established a mapping between the names of packages in the selected distributions.
As \arch was found to have the lowest number of packages (from $\approx$ 7k in 2015 to $\approx$ 10.5k by 2020), we mapped the names of packages found in \arch to those found in other distributions. Manually looking at all possible pairs of distinct package names would have been overly time-consuming (there are up to 62k packages in recent releases of \ubuntu and \debian), we thus first computed the normalised Levenshtein edit distance between each pair of package names, only retaining as mapping candidates those whose distance was below 0.25 from each other.
The mapping candidates were then grouped into sets of package names by agglomerative clustering.
Finally, we manually examined all remaining 6,639 sets of names to determine which of them actually correspond to the same package.
After this mapping process, we had identified 1,065 packages common to all our distributions.

The versioning schemes of 124 of those packages could not be automatically compared between distributions. For instance, the versions of package {\sf lua-socket} are presented in the form of version numbers in most distributions, but as dates in Arch. With no available data source allowing us to determine equivalence between versioning schemes, we had to exclude these packages.
To compare the versions of packages in the distributions, we used \textsf{libversion}\footnote{https://github.com/repology/libversion},
a library that takes into account common versioning notation and keywords. 
For instance, {\sf libversion} would order the following version numbers as such: $1.0rc1 < 1.0 < 1.0patch1 < 1.1$.
We finally excluded 51 packages for which a single version appears in all the selected distributions over the observation period, as it is not meaningful to compare the freshness of such packages. The final dataset contains 890 packages. As these packages are common to the distributions, we avoid distribution-specific packages and cover widely-used packages, including, for instance \textsf{gcc}, \textsf{zsh}, \textsf{nodejs} and \textsf{gimp}.

Since we aim to measure the freshness of packages within \linux distributions, we need to be able to determine when a package version was released by its maintainers. We call this the \textit{upstream release date}, as opposed to the dates at which the package version is deployed {\em downstream} into various \linux distributions. Unfortunately, we found no aggregate source of information on upstream release dates. Using the official sites of the packages themselves is impractical, as these are all formatted differently, potentially requiring a custom script for each package to extract the data. 
As a result, we decided to use the date of first appearance of package versions in one of our selected distributions as a proxy for the upstream release dates.
We will somewhat abusively refer to the proxy as {\em upstream} in the rest of this paper.
Even though our analyses focus on the period $[2015, 2020[$, we applied this procedure to all package versions between 2010 and 2020. This allowed us to approximate the upstream release date of package versions that were already available prior to 2015-01-01.

\section{Quantitative Analysis of Linux distributions}
\label{findings}
\subsection*{$RQ_1$: How prevalent are outdated package versions in \linux distributions?}
\begin{figure}[!ht]
	\centering
	\begin{tabular}{c}
		\includegraphics[width=\figsize]{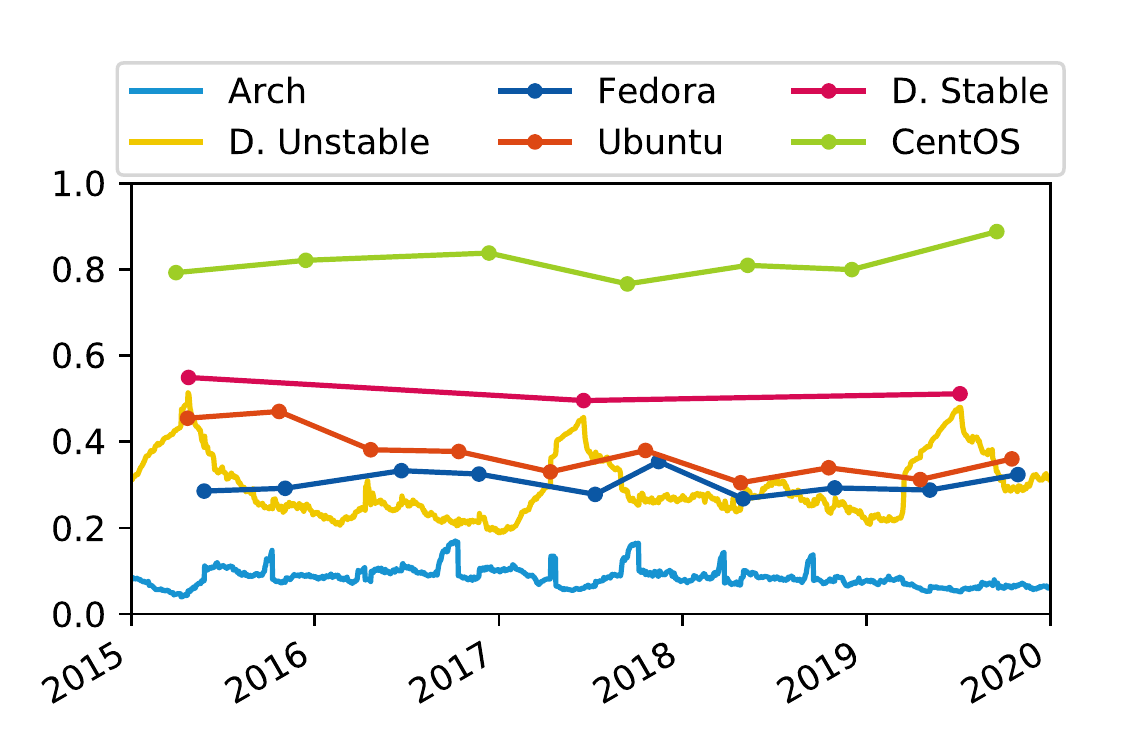}
	\end{tabular}
	\caption{Evolution of the proportion of outdated package versions.}
	\label{outdatedProportion}
\end{figure}

This research question aims to assess the prevalence of outdated package versions within \linux distributions.
A version of a package in a distribution snapshot is \emph{outdated} if a more recent version is present in another distribution on the same date.
Fig.~\ref{outdatedProportion} shows the evolution of the proportion of outdated package versions in the considered Linux distributions.

The proportion of outdated package versions in \arch is very low, oscillating between 5\% and 17\%. 
In contrast, 78\% to 87\% of the package versions in \centosseven are outdated. The proportion of outdated package versions in \stable is around 50\%.
The proportion in \unstable fluctuates wildly between 18\% and 50\%, peaking every 2 years. According to the \debian maintainers, this is due to the adopted release management policy: ``{\em When a Testing release becomes `frozen', Unstable tends to partially freeze as well. This is because developers are reluctant to upload radically new software to Unstable, in case the frozen software in Testing needs minor updates and to fix release critical bugs which keep Testing from becoming Stable}''.\footnote{https://www.debian.org/doc/manuals/debian-faq/ftparchives.en.html\#frozen} So, for a period of six months occurring roughly every year and a half, package versions in \unstable become increasingly outdated, converging towards \stable. Nevertheless, even at the points of the cycle when the fewest packages are outdated, \unstable is more similar to \fedora and \ubuntu than to \arch, which is unexpected for a distribution that serves as a development branch, where one would expect to see recent package versions being made available almost immediately in order to expedite the testing and validation process.
In the case of \fedora, the proportion of outdated package versions hovers around 30\%, peaking at 35\% for \fedora 27.
The proportion in \ubuntu has decreased over the past 5 years, starting at 44\% in \ubuntu 15.04 and reaching 35\% by \ubuntu 19.10.
These observations suggest that not all distributions value keeping packages up-to-date to the same extent.
This is mostly seen by contrasting \arch with distributions that value concerns of stability over freshness, such as \stable and \centos, for which half to three-fourths of the packages are outdated.

\begin{custombox}
	{\bf Findings. }
	The proportion of outdated package versions varies greatly between Linux distributions, from $\sim 10\%$ in \arch to $\sim 80\%$ in \centos.
	Despite being a development distribution, a significant proportion of packages in \unstable are outdated.
\end{custombox}

\subsection*{$RQ_2$: How long have the versions been outdated?}
\begin{figure*}[!ht]
	\centering
	\begin{tabular}{c}
		\includegraphics[width=\figsize*2]{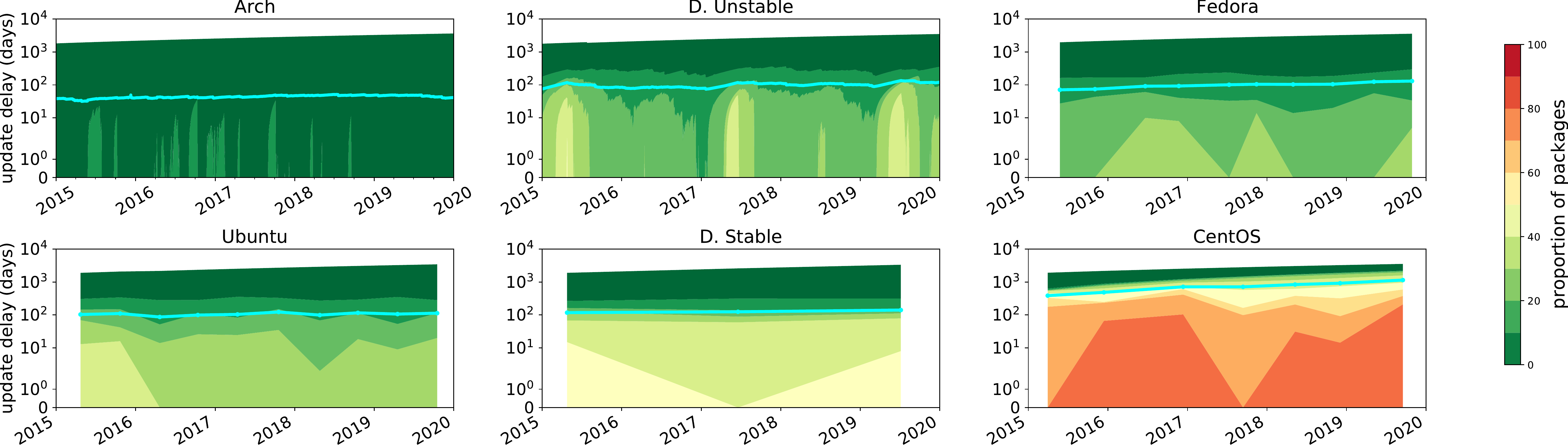}
	\end{tabular}
	\caption{Evolution (on logarithmic y-scale) of the \emph{update delay} for each distribution. Mean update delay is shown \textcolor{visibleBlue}{in blue}.}
	\label{beamUpdateLag}
\end{figure*}

$RQ_1$ established that some distributions use many outdated package versions. However, some versions may have been outdated for a very short time, while others have been outdated for many years.
For instance, {\sf kscreen} is outdated in \ubuntu 19.10 by a single day. Indeed, \ubuntu 19.10 was released on 2019-01-24 with \textsf{kscreen} 5.16.5, but version 5.17.0 was already available on 2019-10-23.
At the other end of the spectrum, \centos 7.6 shipped version 1.5 of package {\sf gzip} even though version 1.6 had already been available for more than five years.
It is therefore important to quantify the time during which distribution maintainers did not seize the opportunity to update.
We thus measure for how long deployed package versions in distributions have been outdated, \ie the time since a more recent upstream version has been available. 
To do so, we define the metric of \textbf{\textit{update delay}} of a package $p$ in a distribution as the time difference between the release date of the distribution containing $p$ and the upstream release date of the first more recent version of $p$. 
If the distribution uses the latest version of $p$, the update delay is 0.
\fig{beamUpdateLag} shows, in increments of 10\%, the \emph{proportion of packages in each distribution having at least a certain value of update delay} (in days) and \textcolor{visibleBlue}{in blue} the evolution of the \emph{mean update delay} (in days) per distribution over the observation period.

We see major differences between distributions. On average, packages in \centos have been outdated by one order of magnitude longer than those in the other distributions. The opposite is observed for \arch, which maintains a mean update delay of less than 52 days.
We see diverging patterns between the \debian family of distributions (\debian and \ubuntu) and the Red Hat family (\fedora and \centos). In the former case, the mean update delay grows relatively little over time, whereas in the latter case, it accrues rapidly over time.
We observe a 87\% increase in mean update delay between \fedora 22 (70 days) and \fedora 31 (131 days). Meanwhile, \ubuntu's mean update delay trends horizontally, allowing it to be lower than \fedora's by the time of \ubuntu 19.04. The mean update delay in \centos almost triples (factor of 2.97) between \centos 7.1 and \centos 7.7, accruing a delay of 761 days over a period of 1,631 days.
For \stable, there is very little increase in mean update delay (16 days only) between release 8 and release 10, even though those releases are 1,533 days apart.
The update delay of \unstable fluctuates as a consequence of the ``partial freeze'' effect prior to \stable releases, as observed in $RQ_1$.

In \arch, the few packages that are sometimes outdated never remain so for long: their update delay rarely exceeds a few tens of days, indicating that maintainers quickly react and update packages to newly available versions. Its rolling release policy facilitates those quick reactions, as there is no need to wait for the next release to update packages.
Few packages are outdated by more than 3 months in \fedora and \ubuntu: 10\% in \fedora and some releases of \ubuntu, 20\% in most releases of \ubuntu.
30\% of \stable packages use versions that have been outdated by 3 months or more. 
\unstable oscillates between update delays that approach \stable around the release of \stable and update delays similar to or even lower than \fedora at other points of the \debian release cycle.
\centos stands in stark contrast with all other distributions, with always over half of its packages outdated by more than a year, and by more than 2 years by the time of \centos 7.6.
The discrepancy we observe between \centos and the other distributions is partly explained by the fact that 41\% of its packages have not been updated over the course of the observation period, despite being outdated from the first snapshot. In most distributions, this phenomenon only concerns $\leq$ 2\% of packages.
For instance, version 0.15.1 of package \textsf{aide} was shipped in \centos 7.7, despite the availability of a more recent version 0.16 on 2013-12-18, 2099 days prior (\ie nearly 6 years)!
In any distribution, amongst outdated package versions, at least 30\% have an update delay of more than half a year and at least 20\% of more than a year.

\begin{custombox}
	{\bf Findings. }
	There is a large discrepancy in update delay between \centos and other distributions: the mean update delay of the first considered \centos release is more than thrice as high as the second distribution (\stable) and ten times as high as the lowest distribution (\arch), and the gap only widens with time.
	Most packages in most distributions have a relatively low update delay, below 3 months. This is not the case for \centos, where half the packages have an update delay of more than a year.
	20\% of outdated packages could have been updated for more than a year in all considered distributions.
\end{custombox}

\subsection*{$RQ_3$: To which extent are deployed packages outdated?}

The \emph{update delay} provided information regarding the time since a package could have been updated to a prior version, but has not been. While informative, this paints an incomplete picture, as it does not measure the amplitude of the outdatedness. For instance, assume that version $v_1$ of package $p$, deployed in distribution $d$ released on date $x$ has an update delay of 2 days. Depending on when $v_1$ was released, $p$ could be missing 2 days' worth of updates or years of updates. The update delay tells us that version $v_2$ was released on day $x-2$. If $v_1$ was released on day $x-3$, then $d$ is only missing 2 days' worth of updates, but if $v_1$ was released on day $x-366$, then it is missing a full year's worth of updates.

\begin{figure*}[!ht]
	\centering
	\begin{tabular}{c}
		\includegraphics[width=\figsize*2]{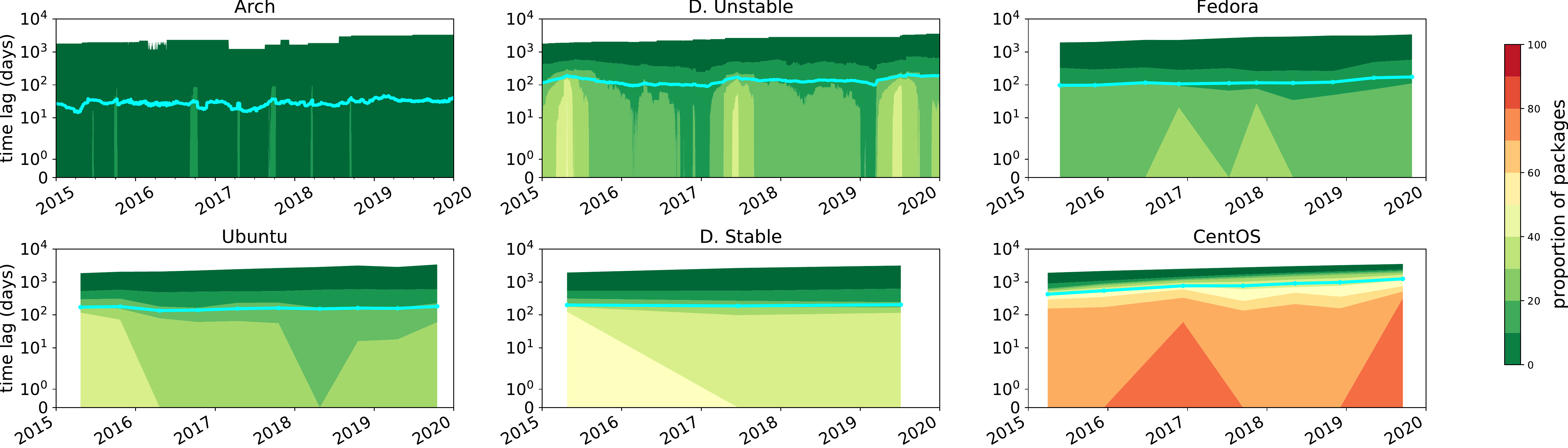} 
	\end{tabular}
	\caption{Evolution (on logarithmic y-scale) of the \emph{time lag} for each distribution. Mean update delay is shown \textcolor{visibleBlue}{in blue}.}
	\label{beamTimeLag}
\end{figure*}

We will quantify the amplitude of outdatedness of package versions in distributions compared to the upstream, by computing the difference between the release date of the most recent upstream version and the release date of the version deployed in a distribution. This corresponds to the \textbf{\emph{time lag}} metric defined in the technical lag framework of Zerouali \etal~\cite{zerouali2019formal}.
For example, \fedora 31 (released on 2019-10-29) contains version 2.0.19 of package \textsf{gob2}. This version was released upstream on 2013-05-07 while version 2.0.20 was released on 2013-12-16. Thus, \textsf{gob2} has a time lag of 223 days in \fedora 31. 
Fig.~\ref{beamTimeLag} shows, in increments of 10\%, the \emph{proportion of packages having this lower threshold of time lag} (in days) and \textcolor{visibleBlue}{in blue} the evolution of the \emph{mean time lag} (in days) per distribution over the observation period.

The mean time lag does not exceed 200 days for most distributions, with the exceptions of \stable (rising to 205 days by \stable 10), and \centos that once again starts more than an order of magnitude higher than \arch (427 versus 28 days) and climbs to over 3 years (1,252 days) by \centos 7.7, more than 6 times higher than the time lag of \stable.
For the most part, most package versions in distributions are less than 3 months older than the latest available version: 50\% in \stable, 60\% in \ubuntu and \unstable, 70\% in \fedora and 90\% in \arch.
In \fedora, fewer than 20\% of packages have a time lag of over 6 months. In \ubuntu, fewer than 30\% do. Even in \stable, more than 60\% (\stable 8) to 70\% (\stable 9 and 10) of packages are using versions less than 6 months old. As prior, \unstable fluctuates according to the \debian release cycle.
\centos is again the exception: half of its packages are outdated by more than a year and a further 10\% by more than 6 months in \centos 7.1. By the time of \centos 7.7, more than 70\% are outdated by more than a year and more than 30\% by more than 5 years!
Considering only the outdated packages in any distribution, at least half of them have a time lag of more than six months, and 30\% of $>$ 1 year.

\begin{custombox}
	{\bf Findings. }
	Distributions have an average time lag of roughly half a year, with the exceptions of \arch (hovering around a month) and \centos (from 1 to 3 years), at opposite ends of the spectrum.
	The majority of package versions in 5 out of 6 distributions are less than 3 months older than the latest available version.
	At least 30\% of the outdated packages are missing more than a year of updates.
\end{custombox}

The metrics of update delay and time lag are complementary, capturing different facets of outdatedness. Two packages can therefore have the same update delay but a very different time lag.
For example, packages \textsf{exempi} and \textsf{enchant} in \stable 9 both have similar update delays (133 and 127 days, respectively), but a very different time lag (12 days and 2454 days, respectively).
Conversely, packages \textsf{xdg-user-dirs} and \textsf{libmpc} in \centos 7.5 both have a time lag of 1722 days, but a very different update delay (246 days and 1564 days, respectively). 
A high time lag is understandable if the update delay is small, as the distribution maintainers have not had much time to incorporate the new version(s).
A high update delay coupled with a low time lag can indicate that the distribution maintainers deliberately skipped a version that introduced an undesired change.
Whereas if both metrics are high, the distribution might be lacking important bug fixes or features.
At the distribution level, the observations we made are consistent across both metrics.

\begin{custombox}
	{\bf Lessons learned. }
	At package level, \emph{update delay} and \emph{time lag} capture different facets of package outdatedness. Nevertheless, at the level of Linux distributions, both metrics are consistent, since the metrics only show important differences for a minority of packages.
\end{custombox}

\section{Discussion}

\subsection{Importance and impact of package freshness}
A survey of Linux users~\cite{legay2020ICSME} revealed that 75\% of them value package freshness and that this is more prevalent for users of cutting edge distributions like \arch and \fedora than users of reputedly stable distributions like \stable and \centos. Installing outdated packages exposes their users to the risk of known security vulnerabilities (unless the security patches have been backported) and users miss out on both bug fixes and new features. Using an outdated version of package $p_1$ may also prevent the adoption of new versions of package $p_2$ that depend on a fresher version of $p_1$.
The aforementioned survey also revealed that users are inclined to update packages through the official repositories whenever possible. This can be explained by the fact that using official repositories comes with the benefit of knowing that the packages have been tested for bugs, co-installability with other packages, dependency requirements and security vulnerabilities and that the user will be notified when future versions are deployed. 
Yet, packages are not always fresh in the official repositories of the distributions. Even in distributions that prioritise freshness, one can find outdated packages.
For instance, \arch shipped version 4.4.2 of package \textsf{findutils} until 2016-02-27, despite more recent versions being available since at least 2011-11-08 (version 4.5.9, present in \fedora 16). This can have several causes, such as the maintainers being reluctant to update to a version that causes excessive breaking changes or incompatibility of the package with other components of the distribution.
Whenever a package is not fresh in (or absent from) the official repositories, other means may be used to install the required package version, such as using community repositories (\eg \textsf{PPAs} for Ubuntu, \textsf{RPM Fusion} for Fedora), third-party package managers (\eg \textsf{Flatpak}, \textsf{Snappy}) or precompiled binaries. 
The survey confirmed that, for instance, 39\% of users resort to binaries to update proprietary software and more than a third use community repositories to install both open-source and proprietary software.
In using these means, users do not profit from the benefits of official repositories. Some, such as installing binaries directly, even carry a risk that the package is malware. 
It is therefore important that packages in the official repositories be fresh.

\subsection{Ranking distributions in order of freshness}

Given the above, we propose to rank distributions in terms of freshness, thereby informing user choice.
To do so, we look at the packages found in a snapshot of each distribution and rank distributions based on the freshness of these packages: the distribution with the highest freshness for a given package receives a position of 1, the one with the lowest freshness a position of 6).
Ties are handled using standard competition ranking.\footnote{https://docs.scipy.org/doc/scipy/reference/generated/scipy.stats.rankdata.html}
We use snapshots of the latest release of point-release distributions. The snapshots used for \arch and \unstable are those on the date of the last point-release snapshot (\fedora). As a result, the snapshots used for most distributions are within two weeks of each other, except \centos (1 month older) and \stable (3 months older).

\begin{figure}[!ht]
	\centering
	\begin{tabular}{c}
		\includegraphics[width=\figsize]{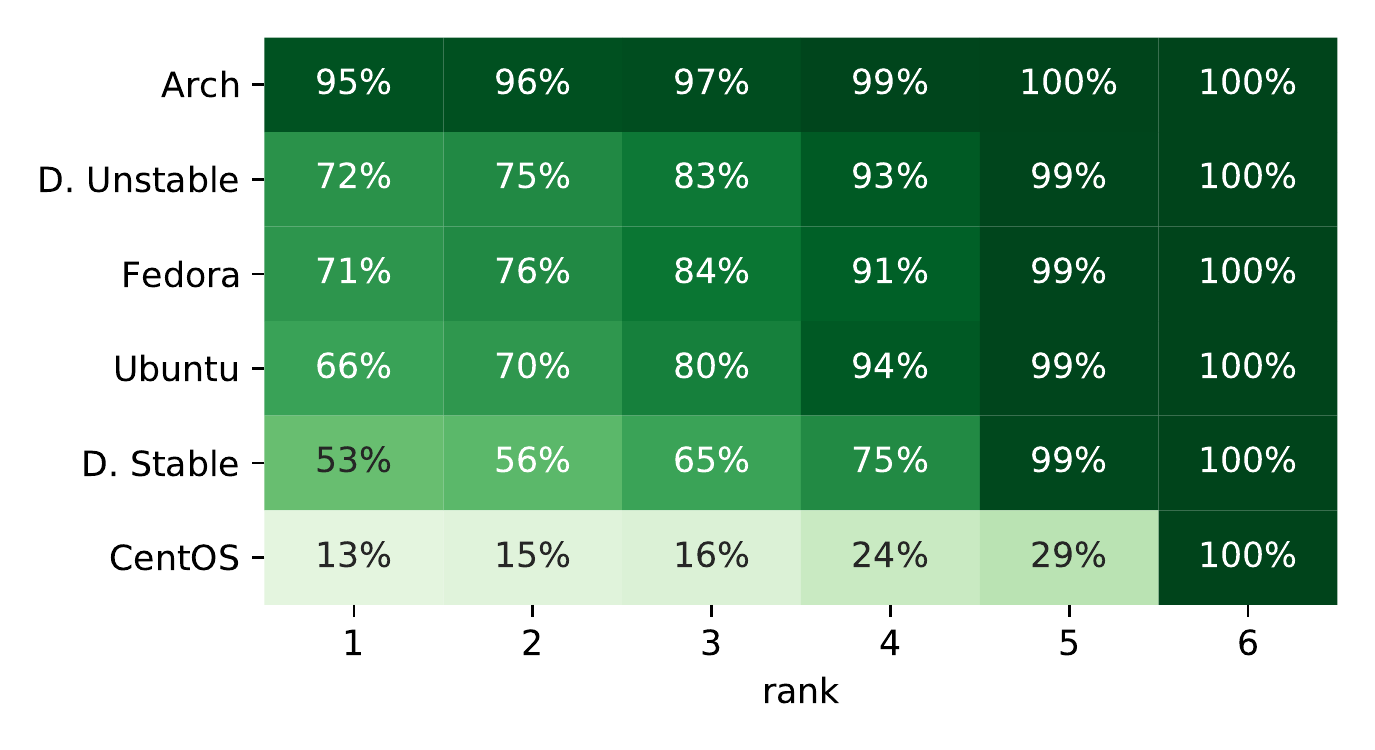}
	\end{tabular}
	\caption{Distribution of time lag rankings for packages in \linux distributions.}
	\label{rankingsHeatmapCumulative}
\end{figure}

We computed the freshness rankings using both update delay and time lag. Since we obtained similar results for both metrics, we only report the time lag ranking. \fig{rankingsHeatmapCumulative} shows, for each distribution, the proportion of packages with regards to their time lag rank.
The numbers shown are cumulative, so a column $n$ shows the proportion of packages ranked from position 1 to $n$. For example, the 76\% value for \fedora in the rank 2 column means that 76\% of its packages have a rank 2 or lower. Since 71\% of its packages have a rank of 1, that means 5\% have a rank of 2.
Five out of six distributions place first more often than not.
This is explained by the fact that from 49\% (\stable) to 93\% (\arch) of packages in the selected snapshot of those distributions are up-to-date, therefore it is expected that many of these packages will tie for first (time lag of 0).
\arch emerges as the clear winner, its packages very rarely being ranked anything but first. Then comes a peloton of 3 distributions: \fedora, \unstable and \ubuntu, ranked first roughly two-thirds of the time. \stable stands slightly back; its packages are ranked first more than half the time (53\%) but lagging behind the packages in the four prior distributions a fourth of the time (25\%). \centos is far behind, its packages being ranked sixth 71\% of the time, which is consistent with the results obtained in prior analyses.

\begin{figure}[t]
	\centering
	\begin{tikzpicture}
	\tikzset{vertex/.style = {shape=ellipse,draw,minimum width=15pt}}
	\tikzset{myptr/.style={decoration={markings,mark=at position 1 with %
				{\arrow[scale=1.5,>=stealth']{>}}} ,postaction={decorate}}}
	\node[vertex] (0) at (0,-1) {Arch};
	\node[vertex] (1) at (2.33,0) {D. Unstable};
	\node[vertex] (2) at (2.33,-1) {Fedora};
	\node[vertex] (3) at (2.33,-2) {Ubuntu};
	\node[vertex] (4) at (4.66,-1) {D. Stable};
	\node[vertex] (5) at (7,-1) {CentOS};
	\draw[myptr] (0) to node[above] {0.24} (1);
	\draw[myptr] (0) to node[below] {0.29} (3);
	\draw[myptr] (0) to node[below] {0.24} (2);
	\draw[myptr] (1) to node[above] {0.14} (4);
	\draw[myptr] (2) to node[below] {0.18} (4);
	\draw[myptr] (3) to node[below] {0.13} (4);
	\draw[myptr] (4) to node[below] {0.13} (5);
	\end{tikzpicture}
	\caption{Order of Linux distributions based on statistical comparison of the time lag of their packages}
	\label{cliffDeltaGraphSparseRearranged}
\end{figure}
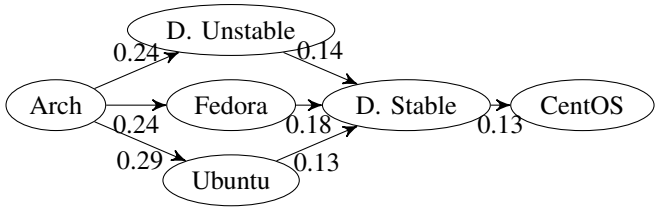

To verify whether there is a significant statistical difference between distributions, we compared the distribution of time lag values between each pair of \linux distributions with a two-sided Mann-Whitney U test~\cite{mann1947}.
For most pairs of \linux distributions, we could reject the hypothesis that the statistical distributions of lag values are identical with statistical significance ($p<0.01$ after Holm-Bonferroni correction~\cite{holm1979simple}), but could not do so for \ubuntu and \fedora, \ubuntu and \unstable and \fedora and \unstable.
We measured the effect size of the difference in time lag between distributions with Cliff's delta $d$~\cite{cliff1993dominance}. \fig{cliffDeltaGraphSparseRearranged} reports the results in a Hasse diagram. An edge from a distribution $D_1$ to $D_2$ indicates that $D_1$ has fresher packages than $D_2$, the reported value is the effect size.
All effect sizes are \emph{small} (and \emph{medium} for $\text{Arch}\rightarrow\text{Ubuntu}$) following the interpretation of Vargha and Delaney~\cite{vargha2000critique}. An order of the relative package freshness of distributions can be inferred, with \arch being the most fresh, followed by the trio \fedora, \ubuntu and \unstable, then \stable, and finally \centos.
Although \unstable comes second with regards to overall package freshness, it fluctuates appreciably in accordance with the release cycle of \stable.

\begin{custombox}
	{\bf Recommendation. }
	Users that place primordial importance on package freshness should default to \arch. As a compromise between package freshness and other factors (such as ease of use or total number of packages), we recommend using \fedora or \ubuntu.
\end{custombox}

\subsection{Comparing \linux user perceptions with quantitative data}

According to our survey~\cite{legay2020ICSME}, \arch users perceived it took on the order of days for upstream versions of packages to be released in the \arch repositories.
We quantitatively confirm that these user perceptions match reality for the majority of packages, as the time lag of 90\% of packages rarely exceeds ten days.
Still, a handful of packages have a significant time lag. For instance, version 1.2.3 of package \textsf{cdrdao} (available since 2010-05-25) is present in \arch until 2018-11-13, even though version 1.2.4 had been available since 2018-07-22, resulting in a time lag of 2980 days (over 8 years).
\fedora and \ubuntu users deemed that it took on the order of weeks. This perception is mostly accurate, as 60\% of packages have a time lag of less than a month in all releases of \fedora and all but 3 releases of \ubuntu (15.04, 15.10, 17.10).
\stable and \centos users considered that it took on the order of months for upstream package versions to reach the downstream official repositories of their distribution. For \stable, that perception is essentially correct: only 30\% to 40\% of packages have a time lag higher than six months.
By contrast, in the case of \centos, it is more appropriate to talk in years as half of its packages have been outdated by over a year in all considered releases.

\begin{custombox}
	{\bf Lesson learned.}
	The user perception of the time it takes for packages to be deployed within their distribution roughly approximates reality, with the exception of \centos users who largely underestimate this time.
\end{custombox}

\section{Threats to validity}
We discuss the threats to the validity of our findings, following the structure established in~\cite{Wohlin2000book}.

\textbf{Construct validity} concerns the appropriateness of using the findings of the experiments undertaken in a study to make inferences, \ie do the experiments measure what they are supposed to.
The principal construct validity threats are related to the use of a proxy (see \ref{methodology}) to approximate the upstream release date of package versions. We did so due to the lack of complete and centralised sources on package version history.
As a result, the release date we consider does not reflect the date when this version was released by the package authors, but the date when this version was first witnessed in one of our distributions. Versions which do not appear in any of the six distributions are not part of the proxy, and those that do appear are assigned a release date that is likely posterior to their actual release date. As such, we potentially underestimated the true update delay, time lag and proportion of outdated packages in the distributions.
As we are principally concerned with comparing the freshness of \linux distributions, underestimating these values is not really an issue, as it applies to all distributions studied.

Additionally, we used snapshots of point-release distributions at the time a new release is made. This does not consider the evolution of the distribution, such as updates to packages in the repository of a distribution made in between two releases. Therefore, for point-release distributions, we only present interpolations of the behaviour between snapshots.

\textbf{Internal validity} concerns the impact of the choices made in carrying out a study on its results.
The use of a proxy can make lag appear more sudden than it really is: a new release of one distribution will make a set of packages of other distributions appear suddenly out of date, when in fact those packages did not all release new versions on the same date. This effect corrects itself whenever those packages are updated. This effect is especially apparent in distributions with a rolling release policy, for which we chose to use daily snapshots. For instance, in \fig{outdatedProportion} we observed that the release of \ubuntu 17.10 on 2017-04-13 caused some packages in \arch to suddenly appear outdated, but this small spike was short-lived as those packages were updated a few days later in \arch. For that reason, we restrained from making observations on local events happening over a few days.

\textbf{Conclusion validity} concerns the reasonableness of the conclusions derived from a study.
There exist two categories of threats to conclusion validity: (a) not detecting a relationship that exist and (b) detecting relationships where there are none. In both cases, the statistical power of the analyses carried out plays an important role.
A potential threat to conclusion validity comes from the relatively small size of our final dataset of 890 packages compared to the total number of packages in some distributions. However, using Cochran's sample size formula, adjusted for a finite population as per ~\cite{israel1992determining}, our sample size of 890 is representative for even the biggest set of packages in our distributions ($\sim 62,000$, in latter snapshots of \stable and \unstable) within a margin of error of 4\% at a confidence interval of 95\%.

\textbf{External validity} concerns the generalisability of the conclusions of a study to a larger scope.
A threat to external validity lies in selecting only packages that are present in all distributions.
This biases the selection in favour of packages that are widespread and generally established in the community over more novel packages. As such, our findings are not generalisable to all other packages.
One could measure the freshness of all packages in all distributions, and compare the freshness of the distributions on this basis. However, such a comparison would be unfair if the set of considered packages is not the same for all distributions.

\section{Conclusion}
\linux distributions rely on external packages to provide their users functionalities that extend beyond those of the \linux kernel. These packages are gathered in the distributions' repositories. Yet, the package versions found in these repositories are not always the latest available ones. A survey of \linux users~\cite{legay2020ICSME} reported that they value package freshness in the official repositories of the distributions they use. In order to allow users to make an informed choice, we therefore assessed the \textit{freshness} of packages in \linux distributions using the complementary metrics of update delay and time lag to measure how up-to-date packages deployed in \linux distributions are compared to the available upstream releases.

We examined the proportion of outdated packages in six \linux distributions, finding a large discrepancy between the most up-to-date distribution (\arch) and the most outdated one (\centos), in terms of the number of outdated packages, update delay and time lag.
As such, users of distributions such as \centos and, to a lesser extent, \stable benefit from new features and bug fixes after other \linux users and might be exposed to known vulnerabilities for a longer period of time. On the other hand, packages being deployed after more extensive testing should better shield those users against stability issues.
From our analyses, \arch emerges as by far the most up-to-date distribution, followed by a trio of distributions (\fedora, \ubuntu and \unstable), then by \stable. \centos is far behind the others. Most packages in \arch, \ubuntu and \fedora are up-to-date, and those that are not are rarely outdated by more than a single version. While \unstable is sometimes as fresh as \fedora and \ubuntu, this highly depends on the \debian release management cycle. 
We therefore would recommend to users who value having fresh packages to choose between \arch, \fedora and \ubuntu.
Surveyed Linux users appear to have an accurate idea of the time it takes to deploy package updates within their distribution, with the exception of \centos users who underestimated that time.

Future work will focus on examining the trade-offs between freshness, security and stability, and ranking distributions according to an aggregate of these criteria.
We also aim to explore the impact of third-party package managers, such as \textsf{Flatpak}, \textsf{Snappy} and \textsf{AppImage} on package freshness: the existence of those package managers grants \linux users another avenue to obtain packages, and therefore potentially mitigate the potentially poor freshness of their chosen distribution. We will examine whether the packages in those package managers are fresh and numerous enough to accomplish this objective.

\section*{Acknowledgment}
This research is supported by the Fonds de la Recherche Scientifique -- FNRS under Grants number O.0157.18F-RG43 (Excellence of Science project SECO-ASSIST) and T.0017.18.
\bibliographystyle{IEEEtran}
\balance
\bibliography{references}

\end{document}